# Simulation and Experimental Analysis of Aerogel's Attenuation for High-Energy Alpha Particles in Fission-Fusion Fragment Rocket Applications

Sandeep Puri*[1], Andrew Gillespie[1], Ian Jones[1], Cuikun Lin[1], Ryan Weed*[2], Robert V. Duncan*[1]

[1]Center for Emerging Energy Sciences, Department of Physics and Astronomy, Texas Tech University, Lubbock, Texas, USA

[2]Positron Dynamics, Livermore, CA

**\* Correspondence:**

sandpuri@ttu.edu; ryan.weed@positrondynamics.com; robert.duncan@ttu.edu;

## 1. Abstract

Emerging studies are geared toward exploring new methods of nuclear rocket propulsion to provide more efficient space transit beyond Earth's orbit. One method is to employ a Fission Fragment Rocket Engine utilizing fissionable layers embedded in a low-density aerogel. A quantitative understanding of particle attenuation is essential for developing a functional prototype that permits fission fragments to escape the layers and contribute to specific impulse rather than being attenuated and generating waste heat. In this investigation, the MCNP code was used to theoretically analyze the attenuation of alpha particles from $^{241}$Am sources within aerogel materials. Simulations were conducted on aerogels with various densities and compositions. These simulations aimed to predict the expected intensity of alpha particles reaching a detector. CR-39 was employed as a Plastic Nuclear Track Detector to assess particle attenuation by the aerogels. The threshold areal density of atoms was found to be around $10^{20}$ atoms/cm$^2$ for the three materials studied in this project. Using a 0.22 mm thick SiO$_2$ aerogel with a density of 90 mg/cm$^3$, which exceeds the threshold, nearly all alpha particles were attenuated. Conversely, employing a 1.6 mm thick graphene aerogel with a density of 12.5 mg/cm$^3$ resulted in an average attenuation of 32.3%.

**Keywords: Aerogel, high energy alpha particle, nuclear rocket, propulsion, plastic nuclear track detector**

## 2. Introduction

A human spaceflight mission poses significant challenges and risks to the crew's health. Hazards like ionizing space radiation, extreme gravitational changes, and prolonged isolation and confinement are among the dangers astronauts encounter[1], [2]. It's crucial to minimize these risks for a successful mission. Enhancing travel efficiency and speed stands as a potential solution to tackle these challenges. This could involve reducing mission duration, which necessitates a rocket equipped with both high thrust (necessary for escaping planetary gravity) and high impulse (essential for achieving rapid interplanetary travel)[3]. Nuclear rocket propulsion offers the promise of greater thrust and higher impulse compared to traditional chemical rockets[4]. This advancement could enable faster, more efficient journeys to distant planets (like Mars) and facilitate the transportation of scientific equipment or materials for use in space far beyond Earth's orbit.

In existing nuclear propulsion models, fission fragments are absorbed within the nuclear fuel, producing heat. Traditional nuclear rocket reactors then create thrust by heating a gas and expelling it via a nozzle[5]. This technique, however, is indirect and not very efficient. Fission fragments, heavy parts of a nuclear core that undergoes a fission reaction, typically carry over 80% of the total released energy. With high atomic mass, a substantial charge state, and extremely high velocity, they're ideal propellants for a magnetic thrust nozzle. Unfortunately, these fission fragments quickly lose speed within solid matter. To make a Fission Fragment Rocket Engine (FFRE) work, the fragments need to be released from a very thin layer or a fuel particle that is much smaller than the attenuation length scale (<10 µm)[6], [7], [8], [9]. These tiny fuel particle elements must also be rigidly contained and accumulated in a very low-density structure to reach a critical mass of fissile fuel, enabling the initiation and sustenance of a nuclear chain reaction. The main challenge in developing a viable FFRE concept lies in meeting both criticality and allowing fission fragment escape. Chapline[10] conceived the FFRE and proposed a solution by rotating micron-thick enriched fuel plates through a moderator structure and extracting the fission fragment via a magnetic yoke. Another concept[11] suggested suspending a cloud of magnetically-confined fissile dust particles within a plasma. However, these ideas faced significant engineering obstacles, including manufacturing fuel elements, thermal and mechanical constraints, and unmanageable launch mass.

Recent breakthroughs in ultra-low-density aerogel materials[12] have introduced the possibility of utilizing a solid substance to capture and assemble these micron-sized fuel particles into a critical assembly. The aerogel's remarkably low density prevents substantial attenuation of fission fragments within the core while also providing effective cooling for the fuel particles through high infrared and visible light transmission. Typically ranging in densities from 0.01 to 400 mg/cm$^3$, aerogels boast extensive total specific surface areas of 30 to 600 m$^2$/g. The discovery of aerogel is attributed to Kistler in 1931, who developed these materials by eliminating the solvent from a liquid component through processes like lyophilization (freeze-drying) or critical point drying. This results in a highly porous, lightweight, and interconnected material[13]. However, a quantitative understanding of the attenuation ratio of aerogel to nuclear particles remains an ongoing area of study.

We propose the use of a Columbia Resin No. 39 (CR-39) was used as a Plastic Nuclear Track Detector (PNTD) to experimentally detect a range of particles with high sensitivity. Capable of effectively detecting protons with energies up to 14 MeV, alpha particles with energies up to 100 MeV, and heavy ions across various energy levels, it serves as a versatile detector. Moreover, it can identify neutrons spanning energies from 500 keV to 20 MeV by observing an enlarged latent damage trail or "track." Details about the discovery and sensitivity of CR-39 detector are present in a number of previous studies[14], [15], [16], [17], [18], [19].

Monte Carlo is a widely adopted statistical technique used for obtaining numerical solutions to physical or mathematical problems, especially when obtaining direct physical measurements proves challenging or unfeasible. In scenarios involving the transport of a large number of charged particles, Monte Carlo relies on random numbers and probability distributions to estimate crucial composite particle characteristics, including energy, position, direction, path-length, and the types of physical interactions experienced within a medium. This technique is capable of determining particle fluence across surfaces or cells and quantifying energy deposition within defined volumes. Numerous Monte Carlo packages such as MCNP[20], FLUKA[21], OPENMC[22], etc. have been developed specifically to address radiation transport problems. Among these, Monte Carlo n-Particle transport code (MCNP 6.2) facilitates the handling of diverse particle transport and interactions across a broad energy spectrum and in three-dimensional arbitrary geometries. It is developed at Los Alamos National Laboratory (LANL), USA and accessible through Oak Ridge National Laboratory (ORNL). It treats particle interactions as either continuous or discrete energies, utilizing cross-section data from the ENDF libraries[23].

In this study, MCNP 6.2 was utilized to theoretically investigate the attenuation of alpha particles from $^{241}$Am sources, while CR-39 PNTD was employed for experimental aerogel attenuation analysis. Alpha particles were used as a substitute for fission fragments due to their interactions with the suspending aerogel materials. Upon colliding with atoms from other materials, they lose their kinetic energy in very short distances. This collision causes those atoms to ionize as they yield their electrons to the incident particle[24]. Various densities and thicknesses of aerogel were also examined in the research. The range of energetic particles is defined as the distance in a medium at which it loses most of its energy due to collisions with electrons. This distance depends on the particle type, particle energy, and the material it travels through. Fenyves and Haiman found that the range of alpha particles for energies between 4-7 MeV can be calculated using **equation 1**[25]. For example, a 5.5 MeV alpha particle emitted by Am-241 can travel approximately 3.9 cm in air, whereas the 4.0 MeV alpha particle from $^{232}$Th has a range of about 2.4 cm. This short stopping distance is due to the double-positive charge, the mass, and the speed of alpha particles, which results in their immense capability to ionize air.

$$R_{air}[cm] = 0.3 \, E \, [MeV]^{3/2} \qquad (1)$$

where $R_{air}$ is the stopping distance in units of centimeters and $E$ is the particle energy in units of MeV. This value for 4 MeV alpha particles is quite similar to the one found from calculations using the open-source software called 'Stopping and Range of Ions in Matter' (SRIM) [26], which is 2.53 cm. As the pressure decreases, the range of energetic particles increases as 1/P.

3. **Simulation and experiments**

   3.1 Simulation

   MCNP 6.2 simulations were employed to predict the anticipated alpha particle intensity reaching a detector. Our next phase involves optimizing an aerogel layer, aiming to permit the transmission of alpha particles through the layer with little or no attenuation. Initially, the transmission rates of 5.5-MeV alpha particles through various aerogel materials were evaluated:

silicon dioxide ($SiO_2$), graphite oxide ($C_2O$), and simulated graphene (represented as elemental carbon). The preliminary study focused on determining the alpha particle transmission rates through these aerogel compositions. **Figure 1** illustrates the simulation's basic geometry, involving a surface source, aerogel layer, and a tally zone utilized for assessing the percentage of transmitted alpha particles. Each layer consisted of cylinders of slightly different dimensions. The surface source was given a radius of 2 cm and a height of 1 cm. Each of these dimensions were arbitrarily chosen because the source zone was only defined in order to visualize the positions and double check the geometry within MCNP. Since the front surface was used as the surface source, the thickness of this region is irrelevant. The aerogel and tally zone radii were set to 2.1 cm. These radii were chosen to be slightly larger than the source radius -. The front surface of the source was positioned 1 mm away from the aerogel and the tally zone was separated from the aerogel by 0.1 mm. The source zone and tally zone were filled with argon. All other regions were filled with air. The first set of simulations was performed for an aerogel consisting of $SiO_2$. A set of simple simulations were performed using a 1 mm-thick cylinder of "aerogel" which is just $SiO_2$ in MCNP. There is a small air gap between the source and the aerogel as well as another small air gap between the aerogel and the tally zone. The percentage of alpha particles transmitted through the aerogel were monitored by dividing the "tracks entering" the tally zone by the "tracks entering" the aerogel. Consecutively the F4 tally (track length estimates of the particle flux) within the tally zone were also monitored. At least $10^8$ histories were used for each simulation.

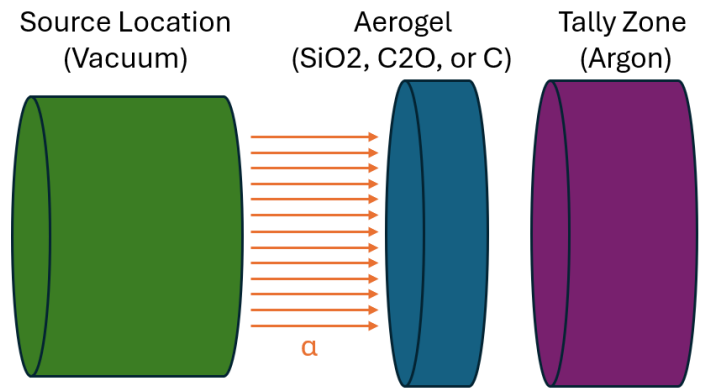

**Figure 1**: The simplified geometry used for simulating alpha particles incident on an aerogel sample of either $SiO_2$, $C_2O$, or C of varied densities.

In order to check the simulation geometry within the MCNP visualizer, each zone was filled with a different material. The material used to display the source location is vacuum space. Since particles originate on the right-most face of this cylinder, the material used for this zone is irrelevant to the simulation. The radius of this cylinder is 2.01 cm. The source definition within MCNP uses the right-most surface and spans a radius of 2.00 cm to avoid potential geometric errors within the simulation. The aerogel is a cylinder filled with either $SiO_2$, $C_2O$, or C. The radius is kept constant at 2.1 cm such that it is slightly larger than the radius of the source. The thickness of the cylinder is set to 1 mm, 2 mm, or 5 mm in these sets of simulations. The density is also varied. The tally zone is a cylinder with a radius of 2.1 cm and a thickness of 5 mm. The material filling the tally zone is argon with a density of 1.664 mg/cm$^3$. This is close to the density of argon at 20 °C and 1 atm[27]. However, this should also be irrelevant to the simulation results presented in this paper because they are only based on the number of particles entering each zone. The rest of the space is filled with air at a density of 1.225 mg/cm$^3$, which is close to the density of air at standard temperature and pressure. The percentage of transmitted particles is based on counting the alpha particles that reach the tally zone and dividing it by the amount that reach the aerogel zone.

### 3.2 Experiment

The experiments utilized silicon (Si) and graphene aerogels, having approximate densities of 90 mg/cm³ and 12.5 mg/cm³, respectively. These tests incorporated a CR-39 PNTD and an $^{241}$Am-point source with a radioactivity level of about 0.88 µCi (activity last measured in August 2023). $^{241}$Am source emits alpha particles with an energy of about 5.5 MeV. Dr. Ryan Weed from Positron Dynamics in San Francisco, CA, provided both the $^{241}$Am-point source and the Si aerogel. The graphene aerogel, labeled G-AREOGEL-75, was acquired from Graphene Supermarket, with an advertised density of around 12.5 mg/cm$^3$ (source: Graphene Supermarket). Additionally, the CR-39 detector used in these experiments was synthesized at TTU.

The alpha particles emitted from a radioactive source encounter a detector with a distributed solid angle. The majority of these particles are emitted at angles lower than 90 degrees. Due to this angle, they traverse longer paths compared to particles emitted directly towards the detector. Consequently, particles at lower angles have more probabilities to interact with absorbing layers and the air between the source and the detector, leading to energy loss[28]. A collimation is needed to minimize uncertainty in particle travel length and air ionization caused by angled alpha particles.

For the experimental design, to achieve this, we fabricated a collimator via 3D-printing with a diameter of 2.54 mm on the top of $^{241}$Am source (Figure S2, supplementary materials). This setup facilitated the positioning of the CR-39 detector, as depicted in **Figures 2(a)** and **(c)**. Different areas of the CR-39 detector were exposed to the americium source for varying durations – specifically, 1, 5, and 10 minutes – to gather data. The experiment comprised four configurations: one without any aerogel layer between the source and the detector (direct exposure), and others with aerogel layers of different compositions, densities, and thicknesses between the source and the detector (as depicted in **Figures 2(b)** and **3(b)**). Subsequent to data collection, the exposed samples underwent etching in a 6.25 molar NaOH solution at 80°C ± 1°C for 30 minutes. The solution was first heated to 80° C using a hot bath before immersing the CR-39 sample. The temperature was continuously monitored with a thermocouple thermometer to maintain the temperature of 80° C within the uncertainty of ±1°C. This etching process enhances particle tracks by attacking and breaking the polymer structure through hydroxide ions, swiftly etching away the majority of the plastic along the paths influenced by alpha particle interactions.

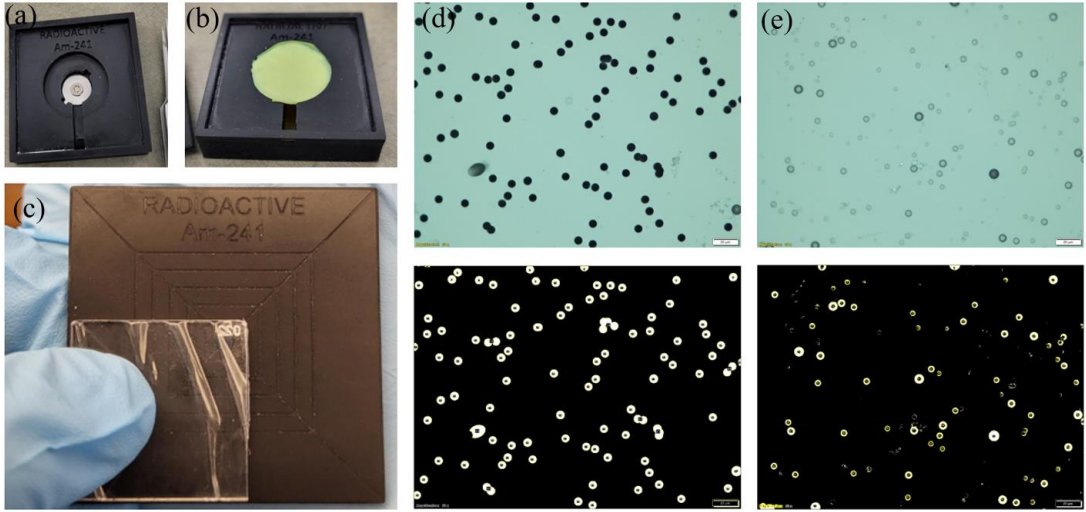

**Figure 2:** (a) A 3D-printed holder housing $^{241}$Am, (b) A 0.22 mm-thick layer of Si-aerogel positioned above $^{241}$Am, and (c) A 3D-printed lid featuring an aperture and a marked scale for reference, facilitating precise exposure control on the CR-39 detector. (d) and (e) The upper panels display images of alpha tracks on the CR-39 and the lower panels display processed images utilizing ImageJ software [29], [30]. (d) Alpha tracks in CR-39 exposed to $^{241}$Am for 1 minute without any intervening aerogel layer. (e) CR-39 exposed to $^{241}$Am for 60 minutes with a 0.22 mm-thick Si-aerogel layer in between.

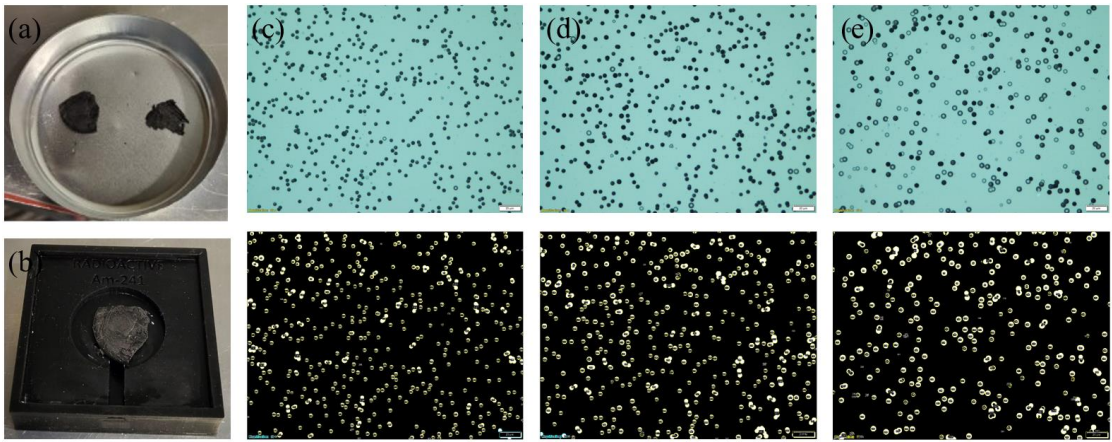

**Figure 3:** (a) Graphene aerogel slices with thicknesses of 1.6 mm (left) and 0.9 mm (right), (b) experimental setup illustrating the positioning of graphene aerogel atop $^{241}$Am, (c-e) alpha tracks in CR-39 exposed to $^{241}$Am for 5 minutes with various intervening layers. (c) Excludes an intervening aerogel layer. (d) When including a 0.9 mm-thick graphene aerogel layer. (e) When including a 1.6 mm-thick graphene aerogel layer.

Following the etching process, the detector samples underwent microscopic examination, and images were captured for detailed analysis and particle estimation in specific regions[31]. These captured images were subsequently processed using "ImageJ" software. Initially, the images were converted into binary format, rendering them as black and white images that highlighted particle tracks as white circular spots, as depicted in the lower panel of **Figures 2**: (d), (e), and **Figure 3**: (c), (d), (e). The software automatically counted these spots, providing an estimation of the number of particles within the designated region. The particle count process using ImageJ is summarized in **Figure S2** included in "supplementary material". To ensure accuracy, each region exposed to the source for a specific duration (e.g., 1 minute) was examined at five distinct sites. The software-derived particle count from each site was averaged to obtain a more precise estimation of the particle number. As When CR-39 was exposed to $^{241}$Am for 1 minute excluding an intervening layer, it resulted in 105 particle track counts, as shown in Figure 2d. When a 0.22-mm-thick $SiO_2$ aerogel intervening layer was included, CR-39 was exposed to $^{241}$Am for 60 minutes, resulting in a particle track count of 72, as shown in **Figure 2e**.

When testing the particle attenuation of a graphene aerogel, each CR-39 slab was exposed to the $^{241}$Am for 5 minutes. When no intervening layer was included, it resulted in a particle count of 484, as shown in Figure 3c. When a 0.9-mm-thick graphene aerogel intervening layer was included, it resulted in a particle track count of 422, as shown in Figure 3d. When a 1.6-mm-thick graphene aerogel intervening layer was included, it resulted in a particle track count of 334, as shown in Figure 3e. A detailed summary of the analyzed data is presented comprehensively in **Table 1**.

4. Results and Discussions

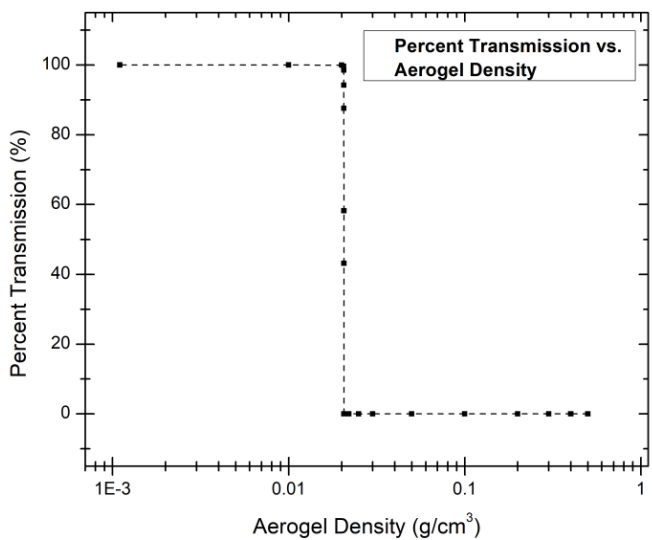

**Figure 4:** The percentage of transmitted 5.5-MeV alpha particles as a function of $SiO_2$ aerogel density.

According to the MCNP simulation outcomes depicted in **Figure 4**, a distinct and very abrupt decrease in alpha particle transmission is observed when utilizing a density around 20 mg/cm$^3$. Nearly all particles are transmitted for aerogel densities below this threshold, while nearly all particles are attenuated for densities above this value. The transition occurs at densities close to 20.6 ± 0.1 mg/cm$^3$, as illustrated in **Figure 4**. Further, the influence of aerogel density on alpha particle attenuation was explored. Initially, our assessment of the aerogel density provided for preliminary experiments was 90 mg/cm$^3$, as provided by the manufacturer. These simulations assumed the aerogel composition to be 100% SiO$_2$. They were straightforward simulations involving a 5.5 MeV alpha particle surface source, an aerogel zone, and a subsequent tally zone. Subsequent simulations were conducted to determine alpha particle transmission rates through aerogels made from alternative materials: Graphite Oxide (C$_2$O) and Graphene Aerogel (C). Interestingly, the maximum density allowing for approximately 100% transmission of alpha particles remained consistent across different aerogel materials. This initial observation aligns with our experimental findings.

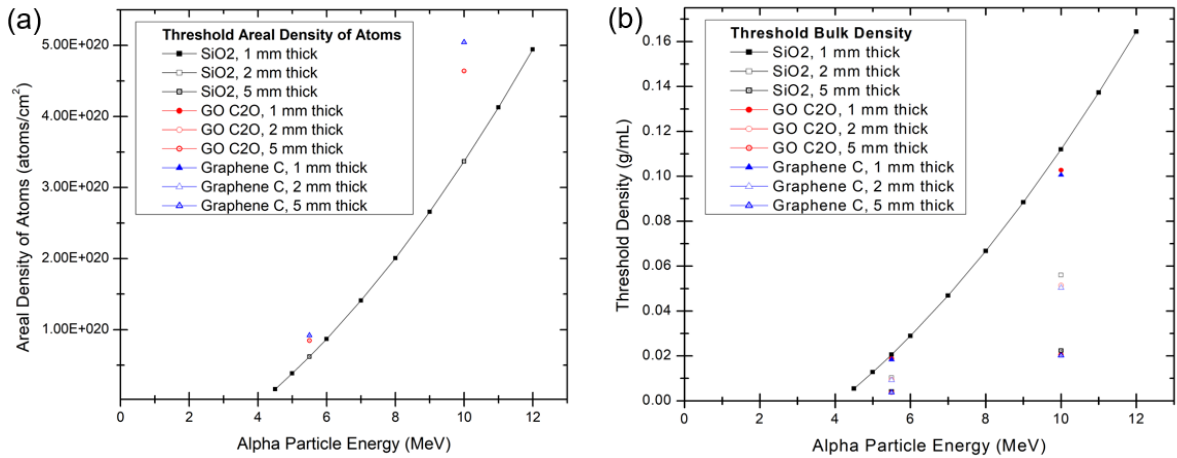

**Figure 5:** *Left*: The threshold areal density of atoms above which aerogels cease permitting alpha particles to be transmitted. *Right*: The threshold bulk density above which aerogels cease permitting alpha particles to be transmitted.

Based on the simulations, the transition from complete to no attenuation of alpha particles occurs precisely at 20 mg/cm$^3$, independent of the aerogel composition, for alpha particles emitted at the $^{241}$Am decay energy (~5 MeV). To delve further into this, simulations were performed with three aerogel materials, varying alpha energy and aerogel thicknesses. When aerogels are below the threshold density, they permit nearly 100% of incident alpha particles to be transmitted through to the following layer. This transmission rate drops drastically for densities above the threshold value. The areal density can be defined as the projection of the attenuating media's individual atom volumes onto the 2D plane perpendicular to the incident particle trajectory. The bulk density is defined as the mass of the aerogel divided by its volume. At a given energy of incident alpha particles, these three materials have a threshold areal density that is similar to one another, as shown in the left panel of **Figure 5**. For 5.5-MeV incident alpha particles, the threshold areal density of atoms is between 6 - 9 x10$^{19}$ atoms/cm$^2$. This threshold areal density of atoms differs more for incident higher energies. For 10-MeV incident alpha particles, the threshold areal density of atoms is between 3 - 5 x10$^{20}$ atoms/cm$^2$.

Since, at a given incident particle energy, the threshold areal density of atoms for one material is the same regardless of the aerogel thickness, it is necessary for the threshold bulk density to vary. If a thicker aerogel is used, then more atoms will occupy each square centimeter in the path of the incident alpha particles. As the thickness increases, it is necessary for the threshold bulk density to decrease, as shown in the right panel of **Figure 5**.

The simulation outcomes align closely with the experimental findings detailed in **Table 1**. Using a 0.22 mm SiO$_2$ aerogel with a density of 90 mg/cm$^3$—exceeding the threshold—nearly all alpha particles are attenuated. Conversely, employing 12.5 mg/cm$^3$ of graphene aerogel results in an average attenuation of ~ 32.3% for a thickness of 1.6 mm and ~ 7.5% for a thickness of 0.9 mm.

**Table 1:** Data collected from the experiment for particle counts for different set up.

| Aerogel thickness & composition (mm) | Exposure time (min) | No. of particles per minute (n/min) | Transmission ratio | Attenuation ratio |
|---|---|---|---|---|
| 0 | 1 | 105.6 | | |
| | 5 | 95.36 | | |
| | 10 | 90.38 | | |
| 0.22 (SiO$_2$) | 1 | 2.0 | 0.019 | 0.981 |
| | 10 | 0.8 | 0.008 | 0.992 |
| | 30 | 0.5 | 0.005 | 0.995 |
| | 60 | 1.2 | 0.011 | 0.989 |
| 0.9 (graphene) | 1 | 98 | 0.93 | 0.07 |
| | 5 | 87.96 | 0.92 | 0.08 |
| 1.6 (graphene) | 1 | 77.2 | 0.73 | 0.27 |
| | 5 | 63.32 | 0.66 | 0.34 |
| | 10 | 57.56 | 0.64 | 0.36 |

At the threshold areal density below which the aerogel becomes transparent to 5.5 MeV alpha particles, around $6.0 \times 10^{19}$ atoms/cm$^2$, the area per atom is $1.7 \times 10^{-20}$ cm$^2$. This implies an average diameter of about 1.3 angstroms (Å) per atom ($1.3 \times 10^{-10}$ m/atom). The typical size of the Si atom is around 2.0 Å and the oxygen atom is around 1.5 Å, but these lengths are estimated for atoms in free space, and this size per atom is dependent on the type of bonding (ionic or covalent) within the crystal in which it is located. Though the size of the SiO$_2$ molecule can span a large range dependent on its crystalline structure or amorphous character, the average diameter per atom at the threshold areal density is close to the diameter of the individual atoms in free space. One would expect these ions to be slightly smaller inside this aerogel structure. It is feasible that the aerogel loses its transparency when the areal density of the atoms fully blocks off any 'free space' between the atoms, as seen by the incoming alpha particle. In the aerogel structure, as in any solid binding of the Si and O atoms, the Si and O atoms have a somewhat positive charge, since they donate electrons to their binding structure. The alpha particle at 5.5 MeV is very small, with a deBroglie wavelength of only 6.1 fm ($6.1 \times 10^{-15}$ m). The frequent grazing collisions do not significantly affect the transport of alpha particles because they result in steady small energy losses as the alpha particles propagate through the aerogel. However, the paths of alpha particles are blocked when they suffer 'hard' collision with the nucleus of the Si or O ion, and experiences most of the core ionic repulsion of the atom's bare nucleus. Though the positively charged alpha particle and atomic Si / O nuclei repel one another slightly in the low areal density limit, when the areal density of the Si / O completely blocks the alpha particle's path, then the aerogel suddenly becomes opaque to the alpha particles. The deBroglie wavelength of the alpha particle decreases inversely proportional to the square root of its kinetic energy, so its wavelength decreases as the particle energy increases. This permits it to 'wedge in' more effectively between the atoms as its energy increases. Further, the stopping power and attenuation of alpha particles in aerogel will likely be different from that in periodic, ordered solid materials, since the aerogel structure is fractal due to its formation through a critical drying process. As such, a direct comparison to predictions of stopping power and attenuation from Bethe – Bloch theory is not feasible. Interesting future work may involve calculations of stopping power and attenuation in a random, fractal distribution of the $SiO_2$ such as in our silicon dioxide aerogel, and explore the possibility that such an analysis leads to an abrupt transition from transparency to absorption at a critical areal density. For alpha particles at 10 MeV, our calculations predict that this sudden transition will occur at an areal density of $3\times10^{20}$ atoms/cm$^2$, which corresponds to a linear dimension of 0.58 Å per atom in a simplified shape assumption of the atoms having closed-packed square cross-sections. But note that, due to this fractal distribution in aerogel, this areal density is as seen by the incoming alpha particle as it projects through the full thickness of the aerogel layer, and not an areal density of atoms in one atomic layer of a regular solid material. Since the alpha particles and the atomic cores within the aerogel are all positively charged, the alpha particle will be electrostatically repelled from a direct core collision until the areal density increases to the point when such avoidance is no longer feasible. At this point, the alpha particle attenuation transitions from zero to nearly complete attenuation. We note that the atomic radii of oxygen is 0.73 Å and of silicon is 1.32 Å,[32] which are close to the 0.58 Å number calculated above. The development of a theory for alpha particle attenuation in fractal materials such as aerogel remains an interesting opportunity for advancement, and our data herein provide are available to test such theories as they emerge. In addition, the diameter of a carbon atom ranges from 1.5 Å in a covalent solid, to 0.6 Å in an ionic solid, so it is smaller than both the Si and O atoms. Hence, one would expect the carbon-rich aerogels ($C_2O$ and graphene) to have a higher threshold aerial density for these 5.5 MeV alpha particles, and indeed they do. This effect becomes much more pronounced at higher alpha particle energies, as the simulations show in **Figure 5**.

In this preliminary study, we have explored alpha particle attenuation in aerogels and the potential application of aerogels to the FFRE concept. Future studies will include the use of actual $^{235}$U fission fragments by aerogel and the viability of creating sub-micron fuel suspended within the aerogels. Additionally, upcoming research will focus on evaluating the degree of heterogeneity of the aerogels, related heating issues, pyrolytic aerogels, and optimizing the magnetic configuration. Small angle x-ray scattering experiments will be used to determine the frequency of densified spots within aerogel samples. Ideally, a subcritical system would be designed with the maximum aerogel thickness that still is transparent to the fission fragments. Materials with a high degree of heterogeneity in their densities will set additional system limitations by attenuating fission fragments, causing local hotspots within the aerogel, and decreasing the thrust efficiency. Designing aerogels with embedded fission fragments much smaller than 1-micron in diameter. A major concern that must be studied is whether it will be physically possible to obtain an adequate density[9]. If not, there may be options involving a subcritical device for power amplification to the fissile isotopes. It is imperative that future experiments include studies of pyrolytic aerogels to contain and explore the viability for the fission fragment concept.

## 5. Conclusions

This research presents a comprehensive investigation into the aerogel's attenuation ratio for high-energy alpha particles, employing advanced MCNP simulation techniques and experimental analyses. Simulations based on silicon, graphene oxide, and pure graphene aerogels with variations in areal density, bulk density, and thickness, demonstrated a transition from complete attenuation to no attenuation at 20 mg/cm$^3$ for alpha particles emitted at the $^{241}$Am decay energy (~ 5 MeV). Further simulations indicated a consistent threshold areal density of atoms regardless of material thickness for a given incident alpha particle energy. Increasing aerogel thickness necessitates a decrease in bulk density to maintain the same threshold areal density. Additionally, the threshold areal density exhibited a slight increase with rising incident alpha energy. The findings from simulations align closely with experimental results, which showed that utilizing 0.22 mm SiO$_2$ aerogels with a density of 90 mg/cm$^3$, which is above threshold, resulted in almost complete alpha particle attenuation. Conversely, the use of 12.5 mg/cm³ graphene aerogel demonstrated an average attenuation of ~ 32.3% at a thickness of 1.6 mm and an average attenuation of ~ 7.5% at a thickness of 0.9 mm. This study has enhanced our understanding of aerogel behavior in alpha particle attenuation, offering valuable insights for diverse applications and help us explore the feasibility of utilizing the aerogels to capture and assemble the micron-sized fuel particles into a critical assembly for our proposed FFRE.

## 6. Declaration

The authors declare that the research was conducted in the absence of any commercial or financial relationships that could be construed as a potential conflict of interest.

## 7. Acknowledgements

This work was supported by NIAC Phase I contract No. 80NSSC23K0592, Department of Energy award No. DE-AR0001736, the Texas Research Incentive Program, and by Texas Tech University. The identification of commercial products, contractors, and suppliers within this article are for informational purposes only, and do not imply endorsement by Texas Tech University, their associates, or their collaborators.

## 8. References

[1] Z. S. Patel *et al.*, "Red risks for a journey to the red planet: The highest priority human health risks for a mission to Mars," *npj Microgravity*, vol. 6, no. 1, p. 33, Nov. 2020, doi: 10.1038/s41526-020-00124-6.

[2] L. J. Abadie, C. Nathan, L. Charles W., S. Mark J., and T. Jennifer L., "The Human Body in Space," *NASA*, 2021, [Online]. Available: https://www.nasa.gov/hrp/bodyinspace

[3] "Ideal Rocket Equation," *NASA*, 2021, [Online]. Available: https://www.grc.nasa.gov/www/k-12/rocket/rktpow.html

[4] R. Raju, "Beyond the Solid Core Nuclear Thermal Rocket: A Computational Investigation into Criticality and Neutronics Performance of Advanced Liquid and Gas Core Reactor Approaches for Next Generation Performance," 2022, doi: 10.7302/6019.

[5] P. R. Mcclure, "Space Nuclear Reactor Development," LA--UR-17-21904, 1345964, Mar. 2017. doi: 10.2172/1345964.

[6] R. Weed, R. V. Duncan, M. Horsley, and G. Chapline, "Radiation characteristics of an aerogel-supported fission fragment rocket engine for crewed interplanetary missions," *Front. Space Technol.*, vol. 4, p. 1197347, Sep. 2023, doi: 10.3389/frspt.2023.1197347.

[7] R. Weed, *Aerogel Core Fission Fragment Rocket Engine*. [Online]. Available: https://www.nasa.gov/general/aerogel-core-fission-fragment-rocket-engine/

[8] R. V. Duncan, C. Lin, A. K. Gillespie, and J. Gahl, "Prospects For A New Light Nuclei, Fission Fusion Energy Cycle," 2023, doi: 10.48550/ARXIV.2308.01434.

[9] J. Gahl, A. K. Gillespie, R. V. Duncan, and C. Lin, "The fission fragment rocket engine for Mars fast transit," *Front. Space Technol.*, vol. 4, p. 1191300, Oct. 2023, doi: 10.3389/frspt.2023.1191300.

[10] G. Chapline, "Fission fragment rocket concept," *Nuclear Instruments and Methods in Physics Research Section A: Accelerators, Spectrometers, Detectors and Associated Equipment*, vol. 271, no. 1, pp. 207–208, Aug. 1988, doi: 10.1016/0168-9002(88)91148-5.

[11] R. Clark and R. Sheldon, "Dusty Plasma Based Fission Fragment Nuclear Reactor," in *41st AIAA/ASME/SAE/ASEE Joint Propulsion Conference & Exhibit*, Tucson, Arizona: American Institute of Aeronautics and Astronautics, Jul. 2005. doi: 10.2514/6.2005-4460.

[12] M. Tabata, I. Adachi, Y. Ishii, H. Kawai, T. Sumiyoshi, and H. Yokogawa, "Development of transparent silica aerogel over a wide range of densities," *Nuclear Instruments and Methods in Physics Research Section A: Accelerators, Spectrometers, Detectors and Associated Equipment*, vol. 623, no. 1, pp. 339–341, Nov. 2010, doi: 10.1016/j.nima.2010.02.241.

[13] S. S. Kistler, "Coherent Expanded Aerogels and Jellies," *Nature*, vol. 127, no. 3211, pp. 741–741, May 1931, doi: 10.1038/127741a0.

[14]     P. J. Potts, "Nuclear techniques for the determination of uranium and thorium and their decay products," in *A Handbook of Silicate Rock Analysis*, Boston, MA: Springer US, 1987, pp. 440–471. doi: 10.1007/978-1-4615-3270-5_13.

[15]     K. O. Inozemtsev and V. V. Kushin, "Comparative analysis of CR-39 sensitivity for different sets of measurable track parameters," *Radiation Measurements*, vol. 91, pp. 44–49, Aug. 2016, doi: 10.1016/j.radmeas.2016.04.010.

[16]     S. Kodaira *et al.*, "A performance test of a new high-surface-quality and high-sensitivity CR-39 plastic nuclear track detector – TechnoTrak," *Nuclear Instruments and Methods in Physics Research Section B: Beam Interactions with Materials and Atoms*, vol. 383, pp. 129–135, Sep. 2016, doi: 10.1016/j.nimb.2016.07.002.

[17]     M. Fujii, R. Yokota, and Y. Atarashi, "New polymeric track detectors of high sensitivity (SR-86)," *International Journal of Radiation Applications and Instrumentation. Part D. Nuclear Tracks and Radiation Measurements*, vol. 15, no. 1–4, pp. 107–110, 1988, doi: 10.1016/1359-0189(88)90110-0.

[18]     V. K. Mandrekar, G. Chourasiya, P. C. Kalsi, S. G. Tilve, and V. S. Nadkarni, "Nuclear track detection using thermoset polycarbonates derived from pentaerythritol," *Nuclear Instruments and Methods in Physics Research Section B: Beam Interactions with Materials and Atoms*, vol. 268, no. 5, pp. 537–542, Mar. 2010, doi: 10.1016/j.nimb.2009.11.015.

[19]     M. A. Rana, "How to achieve precision and reliability in experiments using nuclear track detection technique?," *Nuclear Instruments and Methods in Physics Research Section A: Accelerators, Spectrometers, Detectors and Associated Equipment*, vol. 592, no. 3, pp. 354–360, Jul. 2008, doi: 10.1016/j.nima.2008.04.025.

[20]     "https://mcnp.lanl.gov/; https://silverfirsoftware.com/", [Online]. Available: https://mcnp.lanl.gov/; https://silverfirsoftware.com/

[21]     G. Battistoni *et al.*, "Overview of the FLUKA code," *Annals of Nuclear Energy*, vol. 82, pp. 10–18, Aug. 2015, doi: 10.1016/j.anucene.2014.11.007.

[22]     P. K. Romano, N. E. Horelik, B. R. Herman, A. G. Nelson, B. Forget, and K. Smith, "OpenMC: A state-of-the-art Monte Carlo code for research and development," *Annals of Nuclear Energy*, vol. 82, pp. 90–97, Aug. 2015, doi: 10.1016/j.anucene.2014.07.048.

[23]     M. B. Chadwick *et al.*, "ENDF/B-VII.1 Nuclear Data for Science and Technology: Cross Sections, Covariances, Fission Product Yields and Decay Data," *Nuclear Data Sheets*, vol. 112, no. 12, pp. 2887–2996, Dec. 2011, doi: 10.1016/j.nds.2011.11.002.

[24]     Y. Yan *et al.*, "The differentiation between fission fragments and alpha particles by a Time Projection Chamber," *J. Inst.*, vol. 13, no. 02, pp. P02014–P02014, Feb. 2018, doi: 10.1088/1748-0221/13/02/P02014.

[25]  M. Q.H. and M. H. A., "Alpha-particle Stopping Powers in Air and Argon," *Research & Reviews: Journal of Pure and Applied Physics*, 2017, [Online]. Available: https://www.rroij.com/open-access/alpha-particle-stopping-powers-in-air-and-argon.pdf

[26]  J. F. Ziegler, M. D. Ziegler, and J. P. Biersack, "SRIM – The stopping and range of ions in matter (2010)," *Nuclear Instruments and Methods in Physics Research Section B: Beam Interactions with Materials and Atoms*, vol. 268, no. 11–12, pp. 1818–1823, Jun. 2010, doi: 10.1016/j.nimb.2010.02.091.

[27]  Ch. Tegeler, R. Span, and W. Wagner, "A New Equation of State for Argon Covering the Fluid Region for Temperatures From the Melting Line to 700 K at Pressures up to 1000 MPa," *Journal of Physical and Chemical Reference Data*, vol. 28, no. 3, pp. 779–850, May 1999, doi: 10.1063/1.556037.

[28]  S. Park, S.-W. Kwak, and H.-B. Kang, "High resolution alpha particle spectrometry through collimation," *Nuclear Instruments and Methods in Physics Research Section A: Accelerators, Spectrometers, Detectors and Associated Equipment*, vol. 784, pp. 470–473, Jun. 2015, doi: 10.1016/j.nima.2014.11.045.

[29]  T. Ferreira and W. S. Rasband, *ImageJ User Guide IJ 1.46*. [Online]. Available: imajej.nih.gov/ij/docs/guide/

[30]  C. A. Schneider, W. S. Rasband, and K. W. Eliceiri, "NIH Image to ImageJ: 25 years of image analysis," *Nat Methods*, vol. 9, no. 7, pp. 671–675, Jul. 2012, doi: 10.1038/nmeth.2089.

[31]  Olympus IMS, *Instructions BX53M System Microscope*. [Online]. Available: https://www.olympus-ims.com/en/.downloads/download/?file=285218737&fl=en_US

[32]  A. Kramida and Y. Ralchenko, "NIST Atomic Spectra Database, NIST Standard Reference Database 78." [object Object], 1999. doi: 10.18434/T4W30F.